# Smart Sort: Design and Analysis of a Fast, Efficient and Robust Comparison Based Internal Sort Algorithm

Niraj Kumar Singh[1] and Soubhik Chakraborty[2*]

[1]Department of Computer Science & Engineering, B.I.T. Mesra, Ranchi-835215, India
[2]Department of Applied Mathematics, B.I.T. Mesra, Ranchi-835215, India
*email address of the corresponding author: soubhikc@yahoo.co.in (S. Chakraborty)

**Abstract: Smart Sort algorithm** is a "smart" fusion of heap construction procedures (of Heap sort algorithm) into the conventional "Partition" function (of Quick sort algorithm) resulting in a robust version of Quick sort algorithm. We have also performed empirical analysis of average case behavior of our proposed algorithm along with the necessary theoretical analysis for best and worst cases. Its performance was checked against some standard probability distributions, both uniform and non-uniform, like Binomial, Poisson, Discrete & Continuous Uniform, Exponential, and Standard Normal. The analysis exhibited the desired robustness coupled with excellent performance of our algorithm. Although this paper assumes the static partition ratios, its dynamic version is expected to yield still better results.

## 1 Introduction

The present paper is aimed towards design and analysis of a fast, efficient and robust comparison based hybrid sorting algorithm targeting the universal data set. Here we have proposed a sorting algorithm named 'Smart Sort' which is a robust improvement over the popular Quick sort algorithm. Among the standard sorting algorithms, in the average case (as per mathematical analysis), both the Quick sort and Heap sort are excellent performers. These algorithms exhibit the O ($n\log_2 n$) operation counts for both the operations comparison and assignments. Although excellent for average inputs, the performance of Quick sort degrades to quadratic operation counts for certain data sets, apart from the very well known sorted / almost sorted data sequences. Presence of such patterns has resulted in questioning the robustness of even the average case performance of Quick sort algorithm [12]. A careful analysis of experimental data as contained in [12] and [13] revealed an interesting complementary nature of quick and heap sort algorithms. The quick sort was found to be excellent on continuous data sets, whereas heap sort worked superbly well with the discrete data set. It seems that the productive features of these two standard algorithms if combined together may yield an optimal result. The most natural implication of this fact is to achieve a better hybrid methodology through their proper fusion. With the knowledge of complementary behaviors of these two standard algorithms the next important question in this respect was – 'how to achieve such a fusion?' Natural suggestions include (a) Obtain the balanced partitions entirely through the recursive application of heap functions; (b) Achieve the partitions through embedding the heap properties into the partition function (of Quick sort) appropriately. The first choice was exercised by Singh & Chakraborty in [11]. Here in this paper, we are interested in the second alternative and this is achieved by embedding the heap data structures, both max and min, to achieve a balanced partitioning and hence a robust version of Quick sort algorithm.

Although advanced techniques are available for complexity analysis of algorithms (see [10], [8], [5], and [6]), they often fall flat when it comes to average case analysis This can happen because for a complex code it is difficult to judge which operation is pivotal before applying mathematical expectation and also because the probability distribution over which expectation is taken need not be realistic over the problem domain [1]. Smart Sort yields Ω ($n\log_2 n$) best case count, which is same as that of quick and heap sorts. Although the worst case count of this algorithm is O ($n\log_2^2 n$), the empirical result revealed its conservativeness. The careful embedding of heap routines is an attempt towards avoiding too much random accesses of which heap sort suffers. Also the divide & conquer nature of Smart sort makes it compatible to parallel processing which gives it an edge over the rival heap sort algorithm. Average complexity is explained best by the weight based statistical bounds and their empirical estimates, called empirical O. Smart sort algorithm has average complexity $Y_{avg}(n) = O_{emp}(n\log_2 n)$ which is found to be robust in comparison to Quick sort which lacks this robustness [12]. In our analysis the time of an operation is taken as its weight. The statistical bound estimate is obtained by running computer experiments. A computer experiment is nothing but a series of runs of a code for various inputs.

## 2. The Algorithm: Smart Sort

The Smart Sort algorithm is based on the divide and conquer paradigm. Function "Smart_Partition" is the key sub-routine for this algorithm. The key feature of this procedure is that it performs the "smart" fusion of heap construction procedures (of Heap sort) into the conventional "Partition" function (of Quick sort algorithm) to achieve a reasonably balanced partitioning of the list about the chosen pivot element.

The 'Partition' function used as a sub procedure in 'Smart_Partition' is the standard partitioning function such as one suggested by Hoare [7] (even a minor modification would do). It should be kept in mind that such a 'Partition' function should assumes the first item of the list as the pivot element as this particular choice has the relevance with our algorithm. The functions 'Build_Max_Heap', 'Build_Min_Heap', 'Adjust_Max_Heap', and 'Adjust_Min_Heap' are the standard heap construction procedures (as suggested by Floyd [4]) which are used as sub-procedures with in the 'Smart_Partition' function.

Although an ideal site around which the partition is supposed to be done is the middle position of the list, we have sacrificed this choice at some places. The tradeoffs between the permissible degree of skewness while partitioning the list and heap building costs play an important role in harnessing the true potential of Smart sort algorithm. The idea is to allow the skewness upto certain degree, and to retain the calls to make the partitions essentially of two equal sizes till the time the partition ratio deviates beyond a predefined limit from the desired ideal position ('Mid' in our case). Here the partition ratio is defined as the ratio of lengths of the segments A [Low, J-1] and A [J + 1, High]. The proper selection of Real constants T1 and T2 are crucial to an efficient implementation of this algorithm. As far as the correctness issue is concerned these variables must lie with in the range, $0 \leq [T1, T2] \leq 0.5$. The proposed algorithm is written in a hypothetical language having peculiar resemblance with the much known "C" language.

```
Main_Procedure ( )
{/* A [ ] is the array of elements to be sorted, 'Low' is the position of leftmost element in this list, initialized to 1,
    and 'High' is the right most element of this list and is initialize to n (the number of elements present in the data
    set) */

    Smart_Sort (A, Low, High)
}

Smart_Sort (Real A [ ], Integer Low, Integer High)
 {
     Integer Mid, J, Range
     Real T1, T2

     Mid = (Low + High) /2
     Range = (High – Low) + 1

     IF (Low ≥ High) THEN   RETURN
     J = Smart_Partition (A, Low, High)

     IF ( (J  ≤  Low + T1*Range ) OR  ( J ≥ High – T2*Range )  ) THEN
        Smart_Sort (A, Low, Mid - 1)
        Smart_Sort (A, Mid+2, High)
     ELSE
        Smart_Sort (A, Low, J-1)
        Smart_Sort (A, J+1, High)

} /* end of function Smart_Sort */
```

```
Smart_Partition (Real A [ ], Integer Low, Integer High)
{
    Integer J, Mid, Max, Min, Range
    Real T1, T2

    Mid = (Low + High) / 2
    Range = (High – Low) + 1

    J = Partition (Real A [ ], Integer Low, Integer High)

    IF (J ≤ Low + T1*Range) THEN
    {
        Max = Build_Max_Heap (A, J+1, Mid)
        Min = Build_Min_Heap (A, Mid+1, High)

     IF (Max > Min) THEN
     {
         Exchange (A [J+1], A [Mid+1])
         WHILE (TRUE)
         {
              Max = Adjust_Max_Heap (A, J+1, Mid)
              Min = Adjust_Min_Heap (A, Mid+1, High)

             IF (Max > Min) THEN
                  Exchange (A [J+1], A [Mid+1])
             ELSE
                  Break
         }
     }
     Exchange (A [J+1], A [Mid]) // significant when first element is chosen as pivot

}

 IF (J ≥ High – T2*Range) THEN
 {
    Max = Build_Max_Heap (A, Low, Mid)
    Min = Build_Min_Heap (A, Mid+1, J-1)

    IF (Max > Min) THEN
    {
        Exchange (A [Low], A [Mid+1])
        WHILE (TRUE)
        {
             Max = Adjust_Max_Heap (A, Low, Mid)
             Min = Adjust_Min_Heap (A, Mid+1, J-1)

             IF (Max > Min) THEN
                  Exchange (A [Low], A [Mid+1])
             ELSE
                  Break
         }
      }
      Exchange (A [Low], A [Mid])   // significant when first element is chosen as pivot
  }
```

RETURN (J)
} /* end of the function Smart_Partition */

## 2.1 Description of Smart Sort

Upon execution the procedure 'Partition' returns the index J of element about which the partitioning is supposed to be done. Depending upon the nature (in terms of size ratio) of these two partitions either it is considered to be a reasonably balanced one and no further action (in order to make it even more balanced) is required or this partitioned list is again subjected to further action so as to achieve a better partition through possibly the repetitive application of heap (max/min) procedures. In the first case, the two partitions are A [Low, J-1] and A [J + 1, High], with property that each element in left partition is less than or equal to the pivot element A[J] and each element in the right half is greater than the pivot element. Where as in the later case, upon the termination of 'Smart_Partition' we achieve the sub lists A [Low, Mid - 1] and A [Mid + 2, High]. These partitions obey: MAX {A [Low, Mid]} ≤ MIN {A [Mid + 1, High]}.

## 3. Theoretical Analysis of Smart Sort: Worst and Best Cases

The overall running time of Smart_Sort procedure (in terms of number of comparisons as well as assignments) can be expressed through the recurrence relation: $T(n) = 2T(n/2) + f(n)$, where f(n) is the time incurred in partitioning the list into two reasonably balanced halves. The worst case recurrence is: $T_{worst}(n) = 2T(n/2) + O(n + n\log_2 n)$, which upon solving yielded $T_{worst}(n) = O(n\log_2^2 n)$, whereas the best case recurrence is: $T_{best}(n) = 2T(n/2) + \theta(n)$, which upon solving yielded a running time of $O(n\log_2 n)$. With reference to the discussion carried out in previous section, the existence of conservativeness in this worst case bound cannot be ruled out.

## 4. Average Case Analysis Using Statistical Bound Estimate or Empirical O

This section includes the empirical results performed over Smart Sort algorithm. Average case analysis was done by directly working on program run time to estimate the weight based statistical bound over a finite range by running computer experiments [9] [3]. This estimate is called empirical O [2] [14] [1]. Here time of an operation is taken as its weight. Weighing permits collective consideration of all operations into a conceptual bound which we call a statistical bound in order to distinguish it from the count based mathematical bounds that are operation specific. The credibility of empirical O depends on the design and analysis of our special computer experiment in which time was the response. A detailed discussion can be found in [2].

The observed mean times (in sec) of 100 trials were noted in table (1). Average case analysis was carried out for the thresholds T1=T2=0.01.

*System Specification:* All the computer experiments were carried out using PENTIUM 1600 MHZ processor and 512 MB RAM.

Table 1: Observed mean times in second for Smart Sort algorithm

| Array Size N | Binomial, m=1000, p=0.5 | Poisson, λ=1 | Discrete Uniform, K=1000 | Continuous Uniform, [0-1] | Exponential, mean=1 | Standard Normal, μ=0, σ=1 |
|---|---|---|---|---|---|---|
| 10000 | 0.0121 | 0.0122 | 0.0128 | 0.014 | 0.149 | 0.0148 |
| 20000 | 0.0157 | 0.0159 | 0.0207 | 0.0248 | 0.025 | 0.0297 |
| 30000 | 0.0209 | 0.0289 | 0.036 | 0.0427 | 0.0494 | 0.0426 |
| 40000 | 0.0296 | 0.0377 | 0.0495 | 0.0636 | 0.0567 | 0.0565 |
| 50000 | 0.0368 | 0.0473 | 0.0583 | 0.0764 | 0.0748 | 0.0697 |
| 60000 | 0.04476 | 0.05722 | 0.07864 | 0.0981 | 0.10024 | 0.08322 |
| 70000 | 0.0528 | 0.067 | 0.08566 | 0.11794 | 0.10702 | 0.09792 |
| 80000 | 0.06644 | 0.08288 | 0.0986 | 0.13158 | 0.12214 | 0.1083 |
| **90000** | 0.0706 | 0.08106 | 0.11656 | 0.1413 | 0.13282 | 0.1244 |
| **100000** | 0.08182 | 0.10324 | 0.1352 | 0.14696 | 0.14078 | 0.12192 |

### 4.1 Average Case Complexity for Binomial distribution inputs

The binomial distribution inputs were taken with parameters m and p, where m=1000 and p=0.5 are fixed. The empirical result is shown in fig (1). Experimental result as shown in fig (1) is suggesting a step function that is trying to get close to O(nlogn) complexity. So we can safely conclude that $Y_{avg}(n) = O_{emp}(nlog_2 n)$. The subscript "emp" implies an empirical and hence subjective bound-estimate [11].

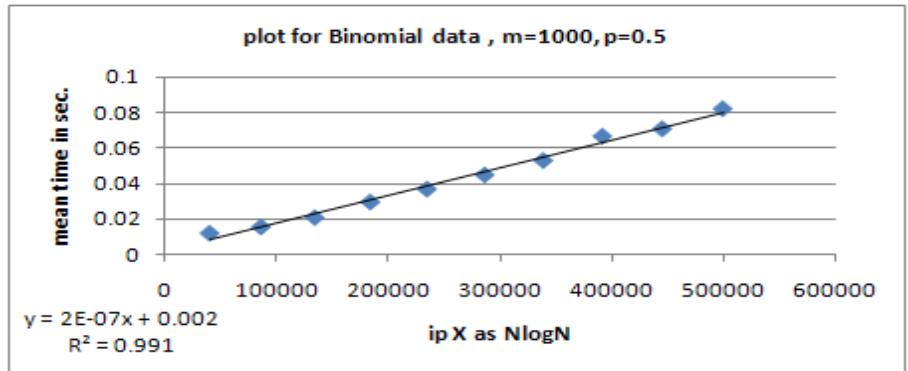

Fig (1): Plot for Binomial distribution data

### 4.2 Average Case Complexity for Poisson distribution inputs

Experimental results as shown in fig (2) is supporting $O(nlog_2 n)$ complexity for Poisson distribution inputs. So we can write $Y_{avg}(n) = O_{emp}(nlog_2 n)$. The input is obtained by putting the value of constant $\lambda =1$.

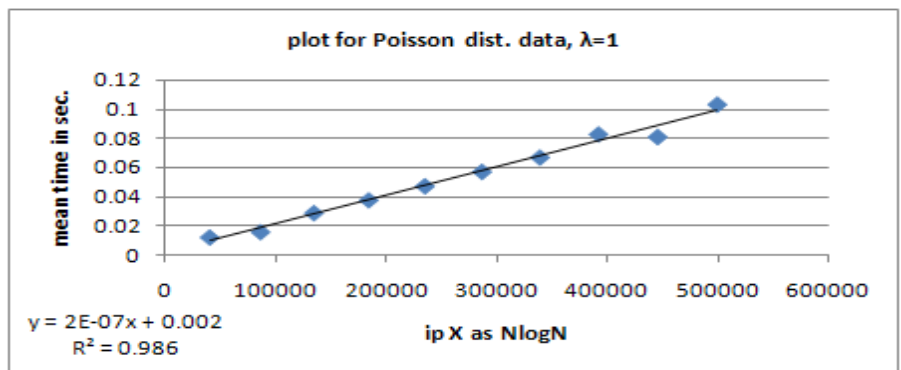

Fig (2): Plot for Poisson distribution data

### 4.3 Average Case Complexity for Discrete Uniform distribution inputs

Experimental result as shown in fig (3) is supporting $O(nlog_2 n)$ complexity for Discrete Uniform Distribution inputs. Looking at the result we are comfortable with the expression $Y_{avg}(n) = O_{emp}(nlog_2 n)$. The parameter k is fixed to k=1000.

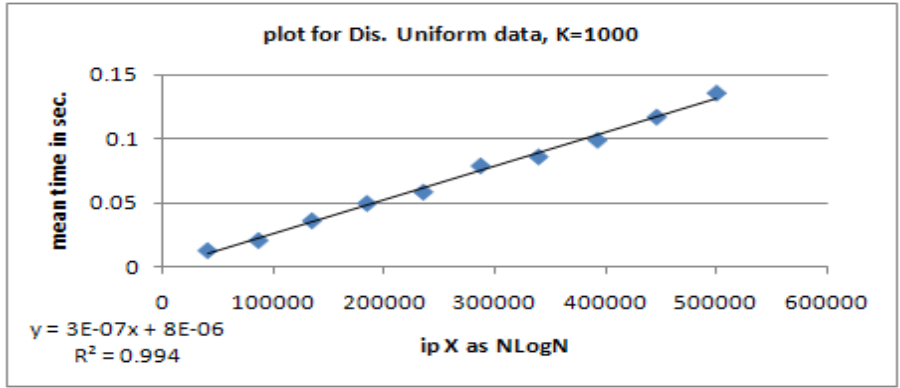
Fig (3): Plot for Discrete Uniform data

### 4.4 Average Case Complexity for Continuous Uniform distribution Inputs

Experimental results as shown in fig (4) is supporting $O(n\log_2 n)$ complexity for Continuous Uniform distribution inputs. So we can again write $Y_{avg}(n) = O_{emp}(n\log_2 n)$. The Continuous Uniform Distribution inputs are taken with parameter mean $\theta$, where $\theta=1$.

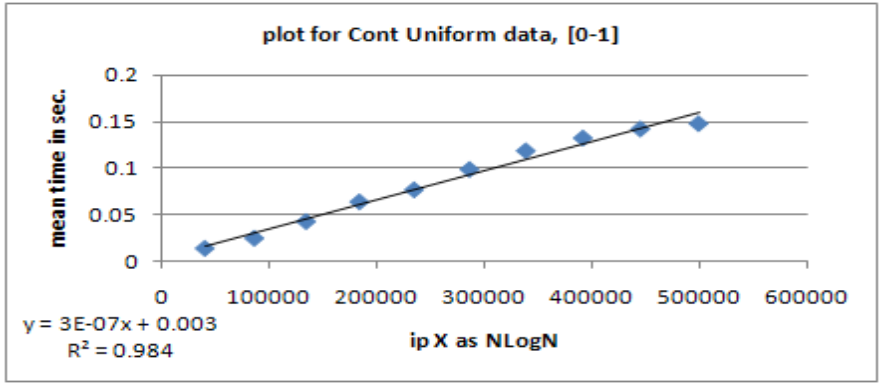
Fig (4): Plot for Continuous Uniform data

### 4.5 Average Case Complexity for Exponential distribution Inputs

Experimental results as shown in fig (5) is fairly supporting $O(n\log_2 n)$ complexity for Exponential distribution inputs. So we can safely put $Y_{avg}(n) = O_{emp}(n\log_2 n)$. The Exponential distribution inputs are taken with parameter mean $\theta$, where $\theta = 1$ is fixed.

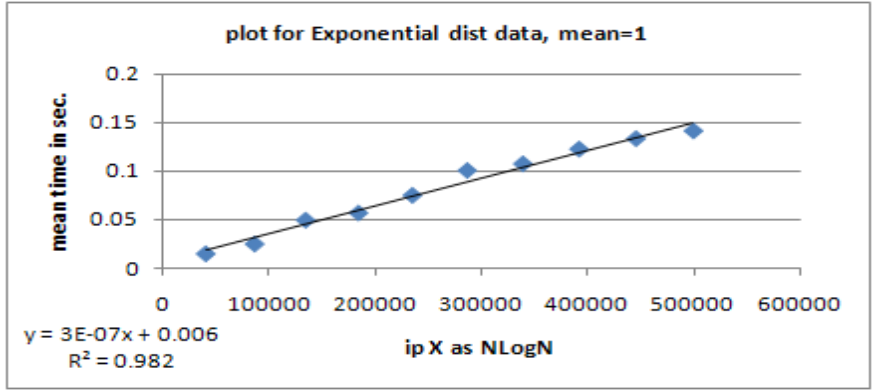
Fig (5): Plot for Exponential data

## 4.6 Average Case Complexity for Standard Normal distribution inputs

Experimental results as shown in fig (6) is supporting $O(n\log_2 n)$ complexity for Standard Normal distribution inputs. So we conclude with $Y_{avg}(n) = O_{emp}(n\log_2 n)$. The parameters taken are mean (μ) & std. deviation (σ), where μ = 0 and σ =1.

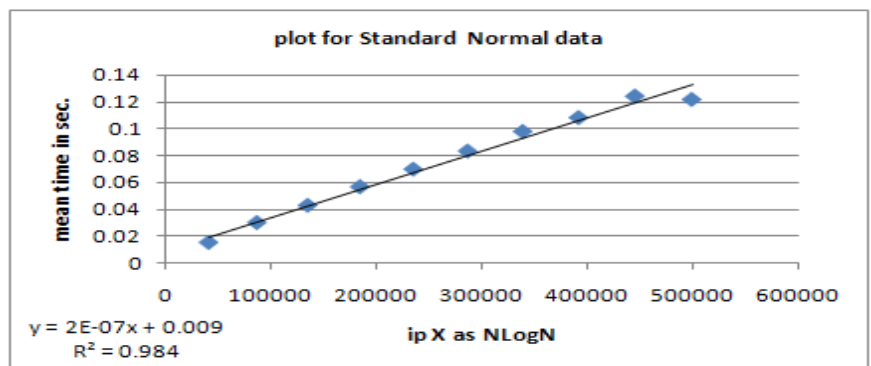

Fig (6): Plot for Standard Normal data

Remark: Empirical O is an estimate and not itself a bound. The term inside it gives the leading term in the empirical model fitted by the statistician to explain and predict time complexity. In this case, since it is obtained by working directly on time, it is estimating a statistical bound.

## 5. Conclusion & Future Works

The experimental results revealed that among the six standard probability distributions examined in this paper, all were supporting the $O_{emp}(n\log_2 n)$ complexity. It guarantees the time bound performance even for the most unfavorable combinations of data, and hence satisfies the robustness criteria. Although the results have been calculated for static partition ratios, we are very sure of the fact that a dynamic choice of such ratios would result in still better performance. And hence we are leaving this task as a rewarding future work. The true potential of an algorithm cannot be realized completely until is subjected to the parameterized complexity analysis, and hence we enlist this task as a related future work.